\documentclass[fleqn,usenatbib]{mnras}

\usepackage{newtxtext,newtxmath}

\usepackage[T1]{fontenc}

\DeclareRobustCommand{\VAN}[3]{#2}
\let\VANthebibliography\thebibliography
\def\thebibliography{\DeclareRobustCommand{\VAN}[3]{##3}\VANthebibliography}

\usepackage{graphicx}	
\usepackage{amsmath}	
\usepackage{xcolor}		
\usepackage{physics}	
\usepackage{multicol}   
\usepackage{xspace}

\title[Intensity Interferometry with H.E.S.S.]{First Intensity Interferometry Measurements with the H.E.S.S. Telescopes}

\author[A. Zmija et al.]{
Andreas Zmija$^{1}$\thanks{E-mail: andi.zmija@fau.de},
Naomi Vogel$^{1}$,
Frederik Wohlleben$^{1,2}$,
Gisela Anton$^{1}$,
Adrian Zink$^{1}$,
Stefan Funk$^{1}$
\\
$^{1}$Erlangen Centre for Astroparticle Physics, Friedrich-Alexander-Universität Erlangen-Nürnberg, Nikolaus-Fiebiger-Str. 2, Erlangen 91058, Germany\\
$^{2}$now at Max-Planck-Institut für Kernphysik, Saupfercheckweg 1, Heidelberg, 69117, Germany\\
}

\date{This is a pre-copyedited, author-produced PDF of an article accepted for publication in Monthly Notices of the Royal Astronomical Society following peer review. The version of record [\cite{10.1093/mnras/stad3676}] is available online at: https://doi.org/10.1093/mnras/stad3676.}

\pubyear{2023}

\begin{document}
\label{firstpage}
\pagerange{\pageref{firstpage}--\pageref{lastpage}}
\maketitle

\begin{abstract}
Intensity interferometry for astrophysical observations has gained increasing interest in the last decade. The method of correlating photon fluxes at different telescopes for high resolution astronomy without access to the phase of the incoming light is insensitive to atmospheric turbulence and doesn't require high-precision optical path control. The necessary large collection areas can be provided by Imaging Atmospheric Cherenkov Telescopes. Implementation of intensity interferometers to existing telescope systems such as VERITAS and MAGIC has proven to be successful for high-resolution imaging of stars. In April 2022 we equipped two telescopes of the H.E.S.S.\ array in Namibia with an intensity interferometry setup to measure southern sky stars and star systems during the bright moon period. We mounted an external optical system to the lid of the telescope cameras, which splits the incoming light and feeds it into two photomultipliers in order to measure the zero-baseline correlation within one telescope in addition to the cross correlation between the telescopes. The optical elements are motorised, which enables live correction of tracking inaccuracies of the telescopes. During the campaign we measured the spatial correlation curves and thereby the angular diameters of $\lambda$ Sco (Shaula) and $\sigma$ Sgr (Nunki), while we also performed systematic studies of our interferometer using the multiple star system of $\alpha$ Cru (Acrux).
\end{abstract}




\section{Introduction}
Hanbury Brown and Twiss (HBT) were the pioneers of astronomical intensity interferometry in the 1960s, when they measured the angular diameter of 32 stars with the Narrabri Stellar Intensity Interferometer \citep{brown1967stellar, HBT_32}. Due to the lack of fast electronics and large area collectors, this technique was subsequently abandoned which led to the development of stellar amplitude interferometry that is applied at instruments, such as the VLTI \citep{haguenauer2010very} and CHARA \citep{CHARA_description}.\\
Although they are highly sensitive, the length of the baseline between the telescopes of these experiments and therefore the optical resolution is limited due to atmospheric turbulence. As a consequence, over the last decade there has been renewed interest in intensity interferometry due to improvements in technology and new large area telescopes in arrays. \citep{Dravins2013} 

While amplitude interferometry employs first order correlation of the electromagnetic wave field at a given spatial distance, intensity interferometry employs second order correlation of the electromagnetic wave field, which is the correlation of photon flux at a given spatial distance. The advantage of this is to make kilometer-scale baselines in the optical band accessible that are possible due to the insensitivity to atmospheric turbulence. Such baselines give rise to angular resolutions of the instrument on the scale of microarcseconds \citep{dravins2016intensity}.

The trade-off is a comparably small signal-to-noise for intensity interferometry observations due to the requirement of two-photon-coincidence measurements \citep{bojer2022quantitative}, which is expressed by the small value of the coherence time (usually on the order of pico- to femtoseconds). Besides the quest for high-time-resolution detectors, one would like to collect as many photons as possible, which requires large light buckets.

Arrays of Imaging Atmospheric Cherenkov Telescopes (IACTs), generally used as gamma-ray observatories, are perfectly suitable for intensity interferometry implementations: the telescopes are amongst the largest that exist on the planet which can focus optical light, observatories typically consist of multiple telescopes separated by baselines on the order of $100\,$m, and the detection mode doesn't allow gamma-ray observations during bright moon times, so that intensity interferometry measurements can be carried out during that time.

Two IACT observatories are already successfully performing intensity interferometry observations. In December 2019, \citep{abeysekara2020demonstration} measured the angular diameters of the two stars $\beta$ CMa and $\varepsilon$ Ori using all 4 telescopes of the VERITAS array as a combined intensity interferometer (VSII). Recently, the VSII star survey has expanded also to fainter stars  up to a magnitude of $+4.2$, while the angular diameters of bright stars can be measured with an accuracy of typically $3$--$4\,$per cent and will undergo further improvements \citep{kieda2022performance}. Further, the MAGIC experiment started intensity interferometry observations in April 2019 with measurements of $\varepsilon$ CMa (Adhara) and $\eta$ UMa (Benetnasch) \citep{acciari2020optical}. Due to the possibility to control individual mirror segments ("active mirror control") the telescopes can also deliver multiple sub-telescope baselines. An extension to the Large Sized Telescopes of the Cherenkov Telescope Array, which is currently under construction, is in the planning. \citep{cortina2022first}

In this paper, we present the first intensity interferometry measurements with the H.E.S.S.\ telescopes in Namibia, where we equipped two telescopes with an intensity interferometry setup mounted to the lid of the Cherenkov cameras. The setup features a beamsplitter at each telescope which enables measurements of the zero-baseline correlation. Our optical pass bands, defined by interference filters, only have a width of $2\,$nm, which is challenging for IACT intensity interferometry. Due to the observatory being close to the southern tropic, it provides excellent observing conditions for targets in the southern sky. We observed Shaula ($\lambda$ Sco) and Nunki ($\sigma$ Sgr), as well as the multiple-star-system Acrux ($\alpha$ Cru).

We introduce the concept of stellar intensity interferometry with the important physical quantities in \autoref{sec:observables}. Section \ref{sec:meas_setup} describes the experimental setup, including the motorised optical elements, which can correct for telescope mis-pointing. We explain our measurement procedure from the start of the night to the end in \autoref{sec:meas_prod}, and give a detailed analysis of the obtained data in \autoref{sec:results}, including the final measurement results of the angular diameters of Shaula and Nunki.

\section{Theory}\label{sec:observables}
Intensity interferometry is sensitive to the second order correlation of electromagnetic waves, i.e. photon rates $I$ at time $t$ and position $\mathbfit{R}$.

\begin{equation}
g^{(2)}(\mathbfit{R}, \tau) = \frac{\expval{I(\mathbfit{R}, t) I(\mathbfit{R} + \mathbfit{b}, t + \tau)}}{\expval{I(\mathbfit{R}, t)} \expval{I(\mathbfit{R} + \mathbfit{b}, t + \tau)}}\,,
\end{equation}

\noindent with $\mathbfit{b}$ (= baseline) being the distance vector between two observers (telescopes), $\tau$ the time difference where the correlation is observed, and $\expval{}$ the time average. Since light from thermal/chaotic light sources is known to show photon bunching \citep{Fox2006}, one measures $g^{(2)} > 1$ for comparably small detector separations $\mathbfit{b}$ and time differences $\tau$. The temporal range in which photon bunching occurs is typically described by the coherence time $\tau_c$, often defined as $\tau_c = \lambda_0^2/(c\Delta\lambda)$, with $\lambda_0$ and $\Delta\lambda$ being the central wavelength and width of the observed wavelength spectrum. However, we decide on a more general definition which can be applied to arbitrary wavelength spectra.

\begin{equation}
    \tau_c := \int_{-\infty}^{+\infty} \left(g^{(2)}(\tau) -1\right) \,\,d\tau
\end{equation}

\noindent The value of the observed coherence time at a telescope baseline $b = |\mathbfit{b}|$ is then expected to be

\begin{equation}\label{eq:coherence_time}
    \tau_c (b) = 0.5 \times k_\textnormal{s}(b) k_\textnormal{T} \frac{\lambda^2_0}{c\Delta \lambda}
\end{equation}

\noindent with $c$ being the speed of light and $k_\textnormal{s} (b)$ being the spatial coherence factor, which is $1$ for $b = 0$ and $0 < k_\textnormal{s} < 1$ otherwise. The factor $0.5$ stems from the fact that unpolarized light is observed. $k_\textnormal{T}$ is a correction factor to the traditional definition of the coherence time $\lambda_0^2/(c\Delta\lambda)$ taking the exact spectral transmission profile into account. The relation between the profile and the temporal coherence time is described via a Fourier transform by the Wiener-Khintchine theorem \citep{Mandel1995}.  Numerical computations of the coherence times using the spectral transmission profiles of the two interference filters which are used - measured by the the manufacturer - yield values for $k_\textnormal{T,1} = 0.842$ and $k_\textnormal{T,2} = 0.848$ for perpendicular incident light.\\
The coherence time in Eq.~\ref{eq:coherence_time} is inversely proportional to the bandwidth $\Delta \lambda$ of the observed light, resulting in large coherence times for small bandwidths. However, if the time resolution of the system is significantly larger than the coherence time (which is true for every field application of intensity interferometry), the signal-to-noise of a measurement is in a first order approximation independent of the used bandwidth, since the photon number is increasing at broader bandwidths, compensating the decreasing value of the coherence time. Nevertheless, restricting the bandwidth by using optical filters in the system is recommended to limit the photon flux on the detectors and avoid systematic error contributions at very small coherence.
The spatial geometry information of the source is encoded in the spatial coherence factor $k_\textnormal{s}(b)$. The degree of coherence - according to the van Cittert-Zernike theorem - is related to the intensity profile of the source via a Fourier transform \citep{Mandel1995}, $k_\textnormal{s}(b)$ can be identified with its square. Hence, significant spatial coherence is only observed within a limited range of baselines $b$, depending on the angular size of the observed object. Large angular diameter stars show spatial coherence of the photons in only a very short baseline range, whereas the spatial coherence of stars of small angular sizes extend over a much longer baseline range.\\

\begin{figure}
\centering
    \includegraphics[width=\columnwidth]{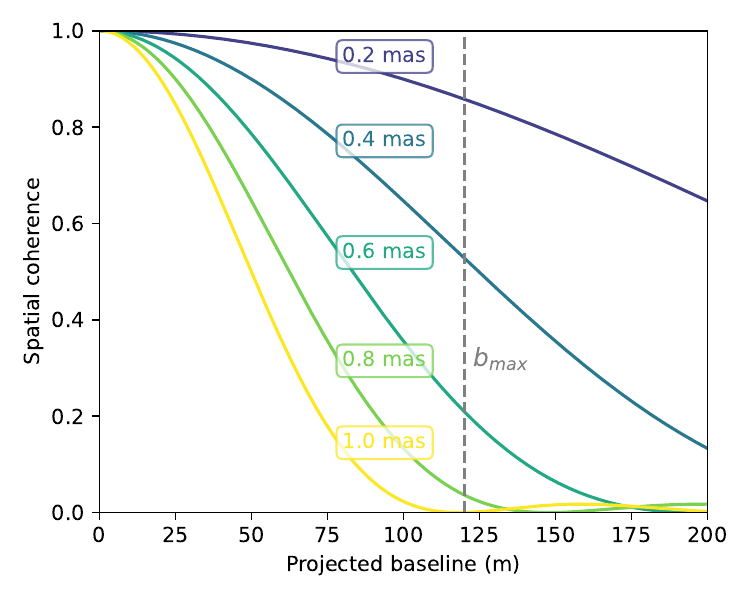}
    \caption{Different spatial coherence curves for different (uniform disc) angular diameters at a wavelength of $470\,$nm. $b_\textnormal{max} = 120\,$m indicates the maximum baseline for the combination of the two used telescopes.}
    \label{fig:scs}
\end{figure}

Fig.~\ref{fig:scs} displays exemplary coherence curves for stars of different angular sizes. Sampling the spatial coherence curve for a star at different baselines allows the angular size of the star to be determined. For the given distance $b_\textnormal{max}$ between two telescopes the star moving along its trajectory in the sky is seen at a range of baselines with the maximum value $b_\textnormal{max}$.

\section{Measurement setup}\label{sec:meas_setup}
\subsection{The H.E.S.S. Telescopes}

\begin{figure}
    \centering
    \includegraphics[width=\columnwidth]{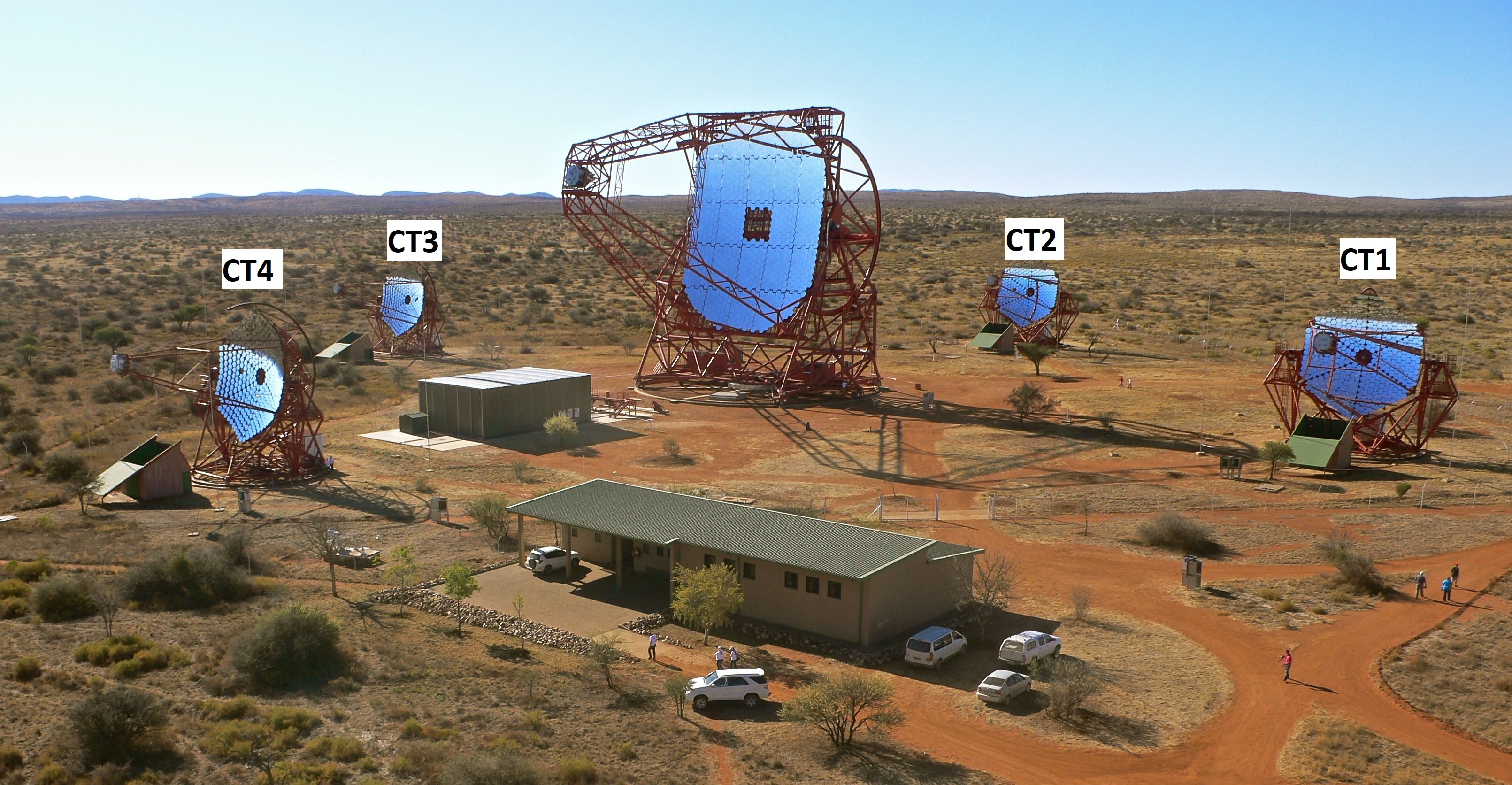}
    \caption{Photograph of the H.E.S.S. site. The inauguration of the Phase I (the four smaller) telescopes was in 2002. Later on, in 2012, the central telescope was added to the project. In 2022, the intensity interferometry setups were installed on telescopes 3 and 4, which are the two left ones in the picture. Part of the data analysis was done in the server room inside of the control building (bottom centre). \protect\citep{HESSpic}}
    \label{fig:hess}
\end{figure}

The intensity interferometry measurements presented in this paper were carried out with two of the H.E.S.S. telescopes (Phase I) located at an altitude of 1835 m.a.s.l. in the Khomas Highlands in Namibia. H.E.S.S. consists of an array of four IACTs arranged in a square with 120\,m side length. At the centre of the array a single 600-m${^2}$ mirror area telescope was added in Phase II in 2012. The phase I telescopes have a diameter of 12\,m and a focal length of $15\,$m. The mirror area is 108\,m${^2}$ consisting of 382 round mirror facets, which are arranged in Davis-Cotton design. A sub-milliarcsecond resolution for intensity interferometry observations is achieved by linking long telescope baselines, large mirror area and advanced optics. \citep{HESStel}\\
A picture of the H.E.S.S. site is shown in Fig.~\ref{fig:hess}.

\subsection{Experimental Setup}\label{sec:exp}

\begin{figure}
    \centering
    \includegraphics[width=0.95\columnwidth]{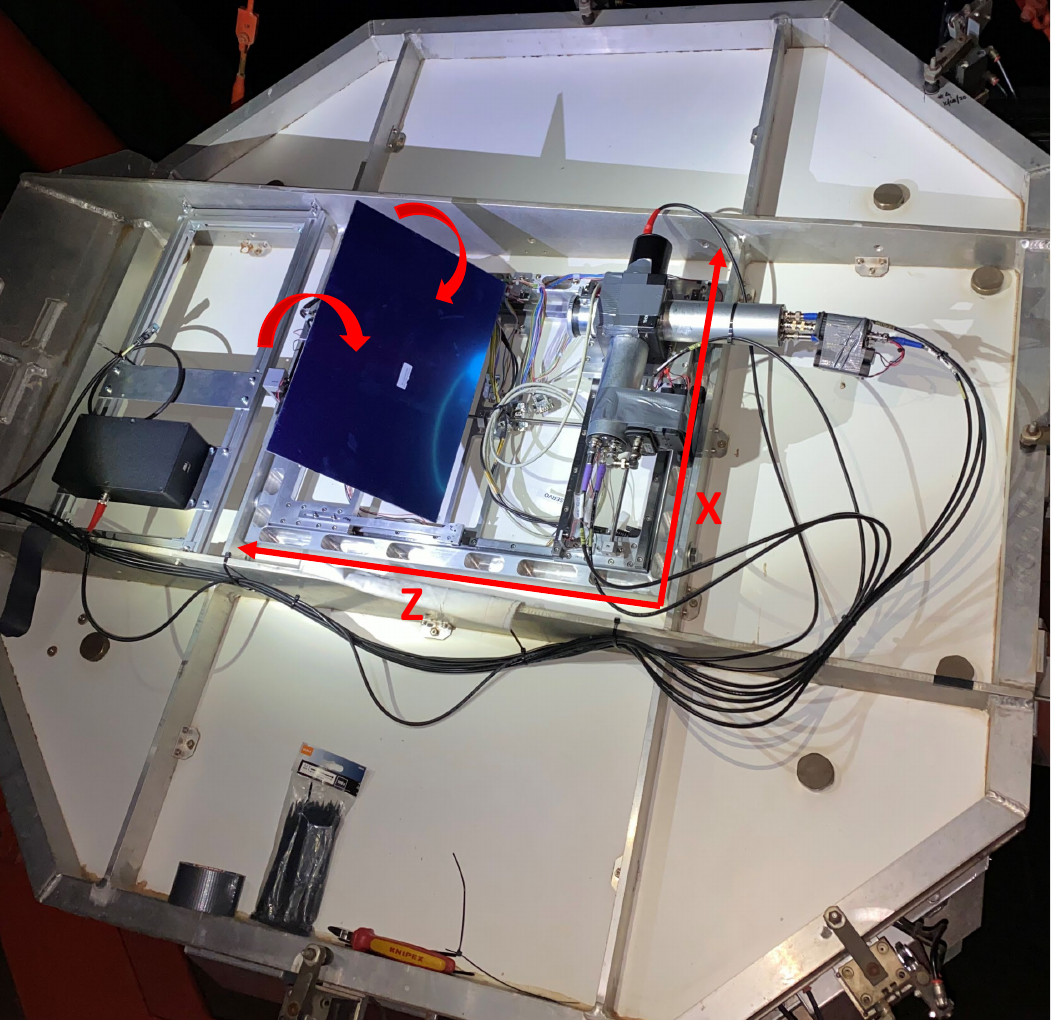}
    \caption{Intensity interferometry setup mounted to the lid of the telescope CT3. As the setup is equipped with motors, the range of motion for each individual component is marked with red arrows.}
    \label{fig:setup}
\end{figure}

\begin{figure}
    \includegraphics[width=\columnwidth]{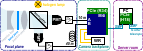}
    \caption{Schematic of the intensity interferometer for one of the telescopes. The setup is identical for each telescope.}
    \label{fig:setup_schematic}
\end{figure}

\begin{figure}
    \includegraphics[width=\columnwidth]{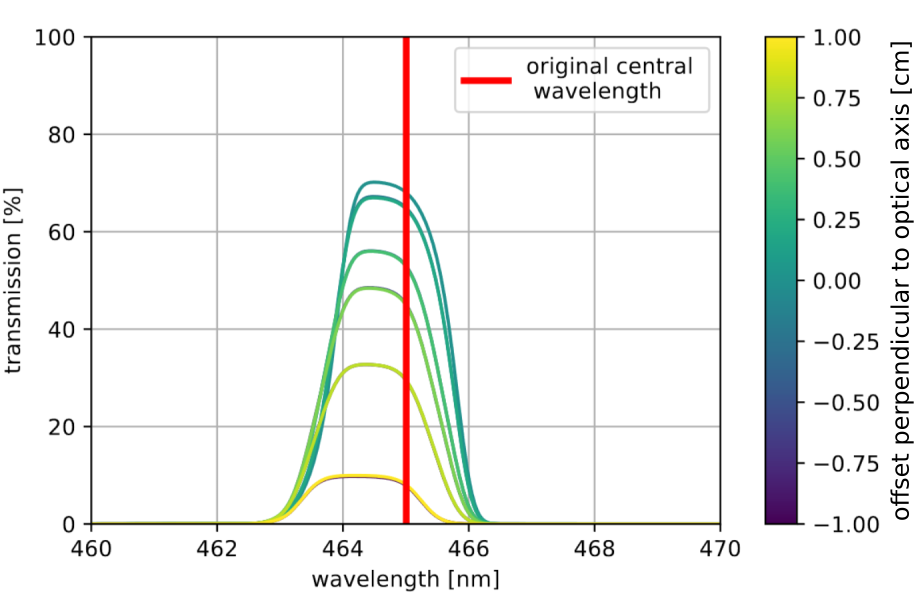}
    \caption{Simulated transmission spectra of the optical filter, taking into account the optical quality of the telescope mirrors. The parallelising lens is shifted by $1.5\,$cm along the optical axis from to the nominal position towards the focal spot. This ensures that most of the light cone is collected by the optical system, while the effect on the transmission spectrum of the filter is kept small. The colors denote different displacements of the tube system perpendicular to the optical axis w.r.t. to the focal spot position. Image adapted from \citep{konrad2021ray}.}
    \label{fig:filter_transmission}
\end{figure}

Our intensity interferometry (II) setup was designed to be mounted onto the lid of the H.E.S.S. Phase I cameras (see Fig.~\ref{fig:setup}). It is a rectangular ($50 \times 63$)-cm${^2}$ structure made of aluminum with bulges to reduce the total weight to 21\,kg with all equipment attached. The setup is mounted at the centre of the lid in the focal spot of the H.E.S.S.\ mirrors. A schematic of the setup can be seen in Fig.~\ref{fig:setup_schematic}. The incident light beam hits the  45$^\circ$ angled mirror (not pictured) which reflects it into the optical path. The light beam, coming from the left side, is divided by a beamsplitter and guided onto the two detectors. The entrance window of the optical system is equipped with a remote controlled shutter to ensure complete darkness if needed. The data is acquired and transmitted to the main workstation where the analysis is performed. \\
The mirror measures ($41 \times 25.5$)\,cm${^2}$ and is 1\,mm thick. It is a SEA-VIS front-surface mirror produced via CNC precision-cut by Pr\"azisions Glas \& Optik GmbH glued to an aluminum frame for stabilisation.\\
The optical path is made of a 2" diameter tube system and consists of a parallelising lens (coated concave lens from \cite{Thorlabs}, $F=-75$\,mm), a LC-HBP465/2-50 narrowband optical (interference) filter (465\,nm CWL, 2\,nm width) and a converging lens (coated bi convex lens from Thorlabs, $F=100$\,mm). The approach of using interference filters of only $2\,$nm bandwidth is challenging, since the light incident on the filter needs to be perpendicular to the filter surface for optimal controlled performance. The telescopes have optical point spread functions on the order of $0.3\,$mrad ($0.5\,$cm) \citep{bernlohr2003optical}. Simulations have shown that the approach works well if the optical system can be adjusted with a precision of a few mm relative to the focal spot position \citep{konrad2021ray}, as shown in Fig.~\ref{fig:filter_transmission}.

After passing the interference filter, the light is split into two beams by the beamsplitter. The beamsplitter is a 50:50 non-polarizing beamsplitter cube from Thorlabs (BS031) with a range of 400--700\,nm.  \\
The detectors are photomultiplier tubes (PMT) by Hamamatsu of type R11265U-300 \citep{Hamamatsu} with a peak sensitivity at 420\,nm and a quantum efficiency of 39 per cent. At our main wavelength of 465\,nm the PMTs still have a quantum efficiency of 30 per cent. The PMTs are powered by a High Voltage (HV) power supply which is located in the back of the H.E.S.S. camera. The first 9 dynodes are powered by the HV whereas the last 4 dynodes have an additional booster power supply to counteract voltage drop. The PMT output is amplified by a factor of $10$. The amplifier is placed directly behind the PMT to minimize electronic noise. The outputs from the amplifiers are then led through an airborne 5 UV shielded polyethylene coaxial cable. Two PMTs are placed at each telescope to also perform zero-baseline correlation measurements. One PMT is connected to a 10-m long signal cable, while the other is connected to a 40-m long cable to avoid crosstalk at the sampling interface. \\
A halogen lamp is installed at the second input of the beam splitter block to function as a light source for calibration runs. The lamp is controlled remotely and its brightness adjusted as needed. \\
The setup is equipped with motors to move each component individually. As shown in Fig.~\ref{fig:setup} the wagon on the right side, where the optics and detectors are attached, can be moved in the camera plane (x- and z-axis). The mirror can be moved up and down in y-axis direction and also around the x- and z-axis. This ensures that the whole light beam is captured by the optical system. When the spot moves out of focus, the components are moved with help of the motors towards the maximum incoming light, thereby ensuring an external online pointing correction.\\
The setup is mounted to the lid of the H.E.S.S.\ camera as shown in Fig.~\ref{fig:setup}. To the left of the setup is a second box for placing the motor control and its power supply, as well as the power supply of the halogen lamp. The HV cables and signal cables are led to the back of the H.E.S.S.\ camera plane where the HV power supply and the digitizer is located. The data is then transferred via optical fibres to the workstation in the control room where data analysis is done after the measurement.

Each of the telescope racks in the backplane of the camera is equipped with a M4i.2211-x8 digitizer from \cite{specint}. It features 8 bit digitization on two input channels, covering possible input voltage ranges between $\pm40$ and $\pm500\,$mV. The maximum sampling speed is $1.25\,$GS/s. For our measurements we used a sampling speed of $625\,$MS/s at a range of $\pm200\,$mV. The digitizers are connected to the workstation's PCIe bus via an Adnacom R34/H18 PCIe-over-fiber extender \citep{pcie}.\\
Synchronization of the two data streams is ensured by connecting both digitizers to the White Rabbit (WR) network of the telescopes. The digitizers are triggered by the pulse-per-second (PPS) of the WR and start to synchronously acquire 2 GS of data, which corresponds to $3.436\,$seconds (note that this is independent of the incoming photon rate since we are sampling the PMT currents). After completion the digitizers wait for the next PPS (which arrives at the 4 seconds mark) to measure the next 2 GS. This results in a duty cycle of $85.9\,$per cent. The procedure was chosen in order to obtain permanent feedback of the functionality of the WR synchronization system.\\
The data is stored to disk during the measurements and correlated offline during the day and after the campaign.

\section{Measurement procedure}\label{sec:meas_prod}
Intensity interferometry observations were performed in April 2022 during the H.E.S.S. moonlight break (full-moon/near-full-moon period) since gamma observations are stopped when the moon has reached 60 per cent of its full moon intensity. Available sources were chosen considering their trajectory, magnitude and configuration. The main observed targets are the stars Shaula ($\lambda$ Sco), Nunki ($\sigma$ Sgr) and Acrux ($\alpha$ Cru). The results are shown in \autoref{sec:results}. \\
First, the telescopes are parked out and slewed to the target of interest for the night, which is then tracked continuously until a new one is chosen or the night is over. While the telescopes are tracking, a PMT calibration measurement with the HV off is performed in order to remove the DC offset later in the analysis. To determine the live photon rates via shape and height distribution of the photon pulses, a PMT pulse calibration measurement with the HV on and low photon rates using the halogen lamp is carried out as well. A detailed description of the calibration procedure is given in \cite{10.1093/mnras/stab3058}.

\subsection{Focusing the targets}
\begin{figure}
    \centering
    \includegraphics[width=\columnwidth]{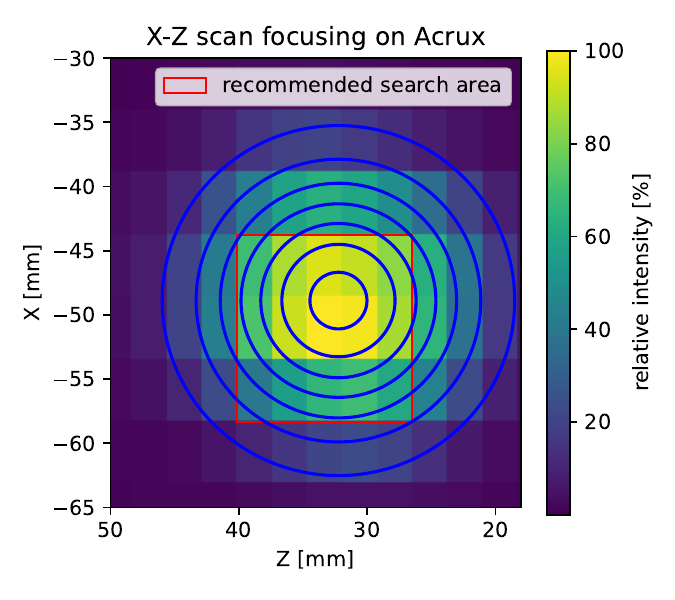}
    \caption{Pointing correction: Heatmap of the recorded photon rates from a scan in the camera plane. The red rectangle shows the area of interest as determined by the first scan. The Gaussian (blue contours) is fitted to the data from the detailed scan. Its centre gives the optimal position for the motors. The definition of the coordinate axes can be found in Fig.~\ref{fig:setup}.}
    \label{fig:scan}
\end{figure}

After the PMT calibration, our optics are focused on the star by optimizing the photon rates using the motors. In a first step both the mirror and optical system are moved in the camera plane (X, Z) and a heatmap of the corresponding rates is produced (compare Fig.~\ref{fig:scan}). The brightest area of the heatmap is then identified as area of interest and further investigated by another (finer) iteration of this procedure. Eventually, the final position of the mirror is determined by fitting a 2d Gaussian to the new heatmap. Furthermore, the optimal distance between the mirror and the optical system is determined by a simple one-dimensional optimization to find the maximum photon rates. The distance between the mirror and the camera plane can be adjusted manually, even though this only yields minor improvements. A rotation of the mirror around any axis did prove to not be necessary.

\subsection{Data acquisition}
\begin{figure}
    \includegraphics[width=\columnwidth]{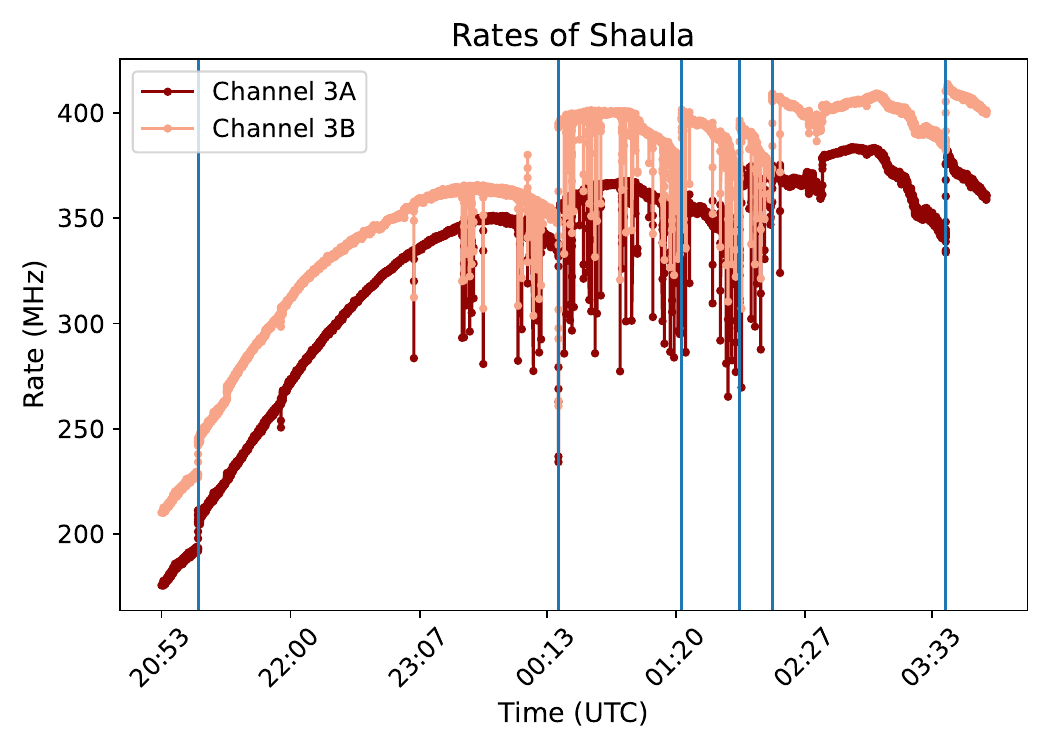}
    \caption{Photon rates of star Shaula in the night of the 19th April 2022. The recorded rates of both detectors of telescope 3 (channel 3A and channel 3B) are shown. As Shaula rises in the beginning of the night the photon rates increase. The vertical blue lines represent some of the times when the motors are refocused to the maximum possible photon rate to counteract rate losses. Due to imperfections in tracking there are frequently occurring dips in the rates, especially between midnight and 2am.}
    \label{fig:rates}
\end{figure}

After having found the optimal motor positions, we start our II measurements while constantly checking the live rates of the stellar target. In Fig.~\ref{fig:rates} an example plot for the rates of Shaula of CT 3 from the night of the 19th April 2022 is shown. Shaula is a star which rises during that time of the year which is shown nicely by the rising rates at the beginning of the night, when the star is increasing in elevation and atmospheric absorption is decreasing. During the middle of the night there are recurring rate losses. This happens when the optical system is not in the focus of the star beam anymore due to the telescope position depending mis-pointing of the telescopes. If the rate loss is on the order of 20 per cent, the mirror is refocused to get back to the maximum rate. This procedure can be seen well at the end of the night at 3:33 UTC time. The large drops in rate at earlier times were recorded while the mirror was refocused. During that process the mirror was driven in and out of focus in different directions until the maximum rate was obtained again. The measurement ends when either the star is less than $20^\circ$ above the horizon which is too low for tracking or the sun is $10^\circ$ below the horizon which is too bright for measurements.

\subsection{Night Sky Background}
\begin{figure}
    \includegraphics[width=\columnwidth]{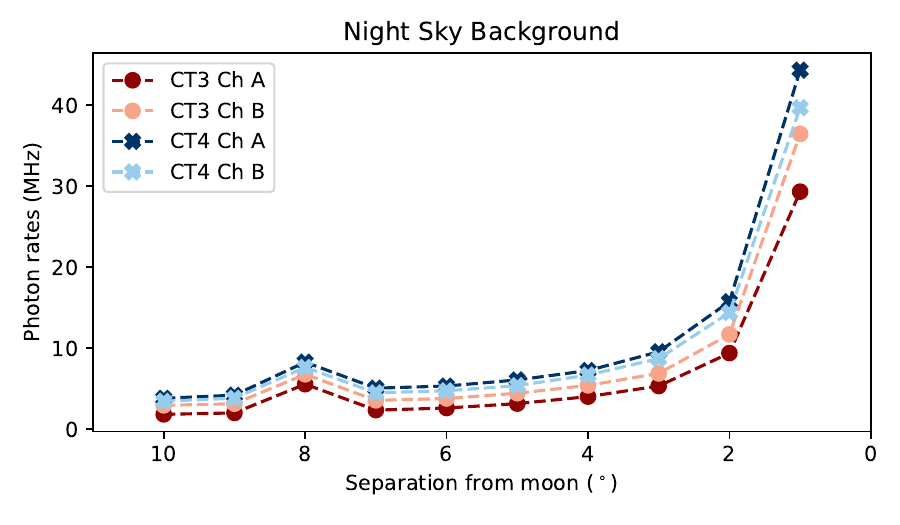}
    \caption{Night sky background at $66\%$ moon phase}
    \label{fig:nsb}
\end{figure}

Uncorrelated background photons from the night sky affect the measurements and have to be taken into account, if the background rates contribute significantly to the total photon rate. We therefore occasionally checked the night sky background, which we expect to be dominated by the moonlight, by slightly mis-pointing the telescopes by not more than $1\,$degree from the target star. In none of these night sky background measurements we observed photon rates which exceeded $10\,$MHz per PMT, which usually is $<5\,$per cent of the photon rates on source, even when the moon was only a few degrees away from the source.
Fig.~\ref{fig:nsb} shows a systematic night sky background measurement with very small separations from the moon, which was as $66\,$per cent brightness in that night. We concluded that night sky background is not a major issue for the bright stars observed in this campaign, as long as the target star is separated by more than only a few degrees from the moon. For fainter stars effects of the relatively higher night sky background rate, such as reduced signal-to-noise and bunching peak heights, need to be considered.

\subsection{End of observations}
At the end of an observation night, the shutter to the optical system is closed, the HV is turned off and the telescopes are parked for daytime. Since the setup is not removed from the lid during the nights where both intensity interferometry and gamma measurements take place, the motors are driven to their parking position. In this position the components, especially the mirror, are stowed in such a way that they don't disturb the opening of the lid for gamma observations. When the telescopes are stowed the amplifiers are turned off and the correlation analysis of the recorded data is started. 

\section{Analysis and Measurement results}\label{sec:results}

\subsection{Correlation procedure}
Two binary files (one from each telescope), each including the two data acquisition channels (see section \ref{sec:exp}) are correlated offline while no measurement is running. The 4 waveform channels are read from the disk and each of the 6 possible correlation channels is calculated, using the correlate function from the python package cupy.
We correlated the measurements of each night during the following daytime. However, since a single $3.436-$seconds measurement takes $13\,$seconds to be analysed using the correlation mechanism of April 2022, we could only analyse a fraction during our time on site. The remaining data was correlated later in the lab once the computers and hard drives arrived back in Erlangen.

\subsection{Noise cuts}
Although the PMTs are encased in aluminum tubes to reduce noise, the Fourier transforms of recorded photomultiplier currents as well as the resulting $g^{(2)}$ functions show remaining sharp peaks at several frequencies. Noise peaks would also show up in measurements carried out in the lab in Erlangen even with further aluminum shielding. Therefore, the data has to be cleaned before further analysis is done. The first cleaning is done with a lowpass filter with a cut-off frequency of $200\,$MHz. It is applied to all 6 correlation channels. For higher frequencies a notch filter is applied.

\subsection{Zero-baseline correlations}\label{sec:zero_baseline}
In both telescopes, the zero-baseline correlation between the two channels in one telescope is calculated. We denote this type of (single telescope) correlation as "auto-correlation", even though this is not in accordance with its usual meaning of correlating a single signal with itself. The purpose of the auto-correlation is two-fold: it provides an additional data point for baseline zero in the spatial coherence plot, and it enables monitoring systematic effects in the telescopes that can affect the correlations. Since the baseline between the two channels in one telescope does not change over time, the measured coherence time should neither. Possible effects including variation of the angle of incidence of the incoming light on the interference filter however can result in a varying coherence time, which can also transfer to the cross correlations between the telescopes, and should be checked carefully.

\begin{figure}
    \includegraphics[width=\columnwidth]{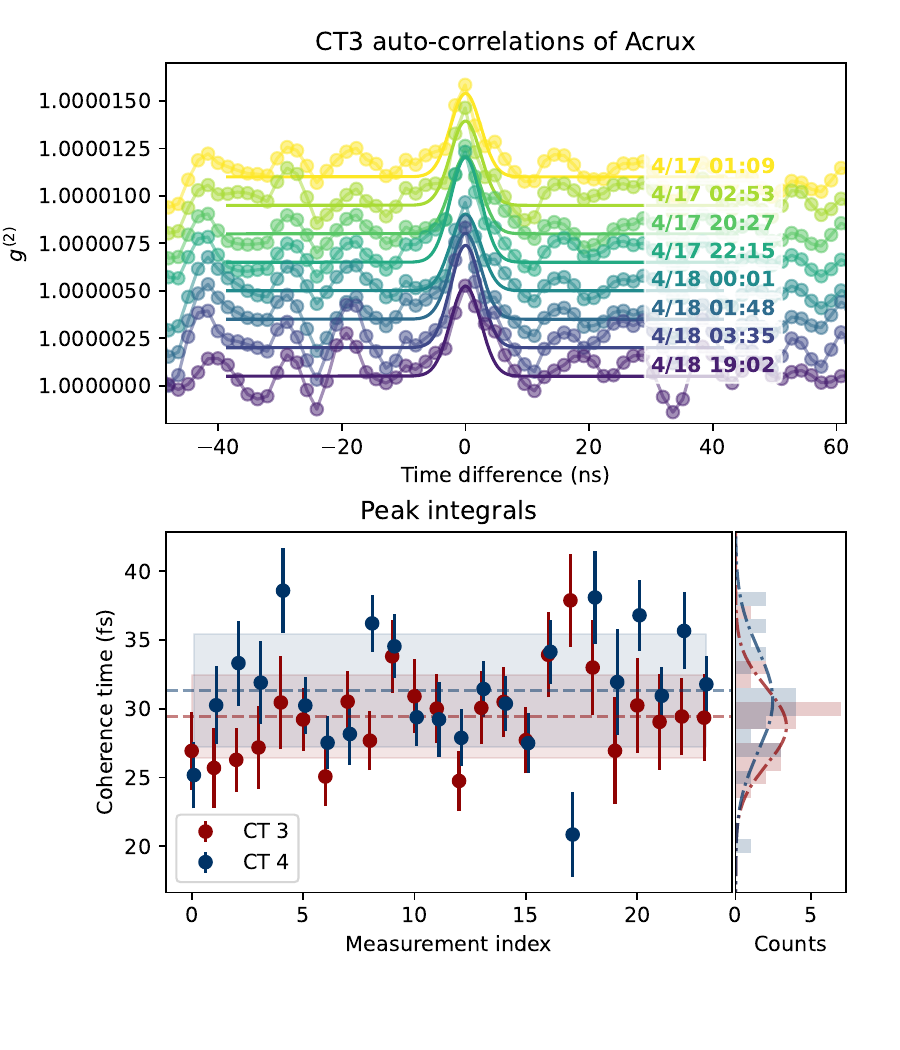}
    \caption{Top: zero baseline ("auto-")correlations of Acrux measured in CT3. For reasons of clarity, only every third measurement segment is plotted, together with a Gaussian fit. The times in the legend correspond to the central time of each segment. Bottom: integrated coherence times at zero baseline of all measurement segments of Acrux in both CT3 and CT4. The shaded areas show the standard deviation of the data around the mean (dashed lines).}
    \label{fig:acrux_ac}
\end{figure}

Fig.~\ref{fig:acrux_ac} (top) shows auto-correlation measurements of the star system Acrux at different moments in time. For a better overview the measurements are plotted with an offset on the vertical axis with respect to each other. The photon bunching peak at time difference $0$ is clearly visible, but also systematic patterns, especially at negative time differences, are apparent. Systematic oscillations of the same magnitude may also be present in the peak region and artificially increase or decrease the found peak integrals. This results in an additional systematic uncertainty, which is of special interest in the investigation of the combined spatial coherence and auto-correlation (figures \ref{fig:nunki_sc}, \ref{fig:shaula_sc}, \ref{fig:acrux_sc}) of the observed targets, and is further discussed in section \ref{results}.

The bunching peak can be described sufficiently well by a Gaussian shape, its integral is expected to be equal to the coherence time in Eq.~\ref{eq:coherence_time} for $k(b=0)=1$ if the star is assumed to not be resolved by the telescope's diameter. We obtain

\begin{equation}
    \tau_c(b=0) = 0.5 \times 0.842 \times \frac{(465\,\textnormal{nm})^2}{c \times 2\,\textnormal{nm}} = 152\,\textnormal{fs}
\end{equation}

The peak integrals of every measurement interval of Acrux, ranging over three different nights, are displayed in Fig.~\ref{fig:acrux_ac} (bottom) for both telescopes. The measured coherence times show a clear deviation from the expected value of $152\,$fs, which directly results in a reduced signal-to-noise. Apart from the fact that the multiple star system might already be partly resolved by a single telescope, this issue is further discussed in section \ref{sec:data_analysis} for Nunki and Shaula, where the same systematic is found for the combined auto- and cross correlation data. The measured coherence times, however, are consistent with each other and constant over time, indicating that the interferometer's and telescope's systematics are stable and do not further affect the measurement procedure.

\subsection{Optical path delay compensation}
\begin{figure}
    \begin{centering}
        \includegraphics[width=0.3\columnwidth]{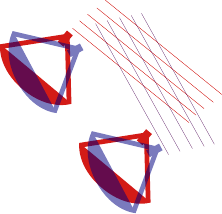}
        \caption{Different optical path delays between both telescopes at different pointing positions. In the red scenario the light arrives at the upper telescope first, while in the blue scenario it is the lower telescope.}  
        \label{fig:path_delay} 
    \end{centering}
\end{figure}

While the telescopes keep tracking the same target in the sky, the optical path length of the light arriving at both telescopes is different and continuously changes during the measurement. Fig.~\ref{fig:path_delay} shows two exemplary cases where different pointing directions lead to the light arriving at different telescopes first.

The optical path delay directly relates to the time position $\tau$ in $g^{(2)}(\tau)$ of the expected correlation peak. Since we need to accumulate data over time spans of order 1 hour to measure significant bunching signals, we need to continuously correct for the optical path delay within one measurement segment to prevent washing out the signal peak over too many time bins.
The correction is applied digitally by checking the absolute time of acquisition of every $3.4-$s file. The position of the observed object in the sky is computed and from that the relative optical path delay and delay time between the two telescopes are calculated. Each elementary $g^{(2)}$ function consisting of $3.436\,$s of data is shifted by the corresponding number of time bins before the averaging process.

\subsection{Cable delay correction}
After the optical path delay correction, the remaining electronic delay - induced by the different cable lengths - results in bunching peaks close to two different time delays: if two channels of similar cable lengths are correlated, the bunching peak is close to $\tau = 0$. That is true for the correlations of CT3 channel A with CT4 channel A, which both use signal cables of $\approx 10\,$m, and the correlation of CT3 channel B with CT4 channel B, which both use signal cables of $\approx 40\,$m. In the case of correlating one $10-$m channel with one $40-$m channel, the additional signal travel time in the longer cable shifts the bunching peaks to $\tau \approx 117\,$ns, corresponding to a signal travel speed in the cables of $0.855\,c$. This is the case for the correlations of CT3 channel A with CT4 channel B and CT4 channel A with CT3 channel B (and also for the auto-correlations in each telescope, which have already been cable-delay corrected in Fig.~\ref{fig:acrux_ac}).

\begin{figure}
    \includegraphics[width=\columnwidth]{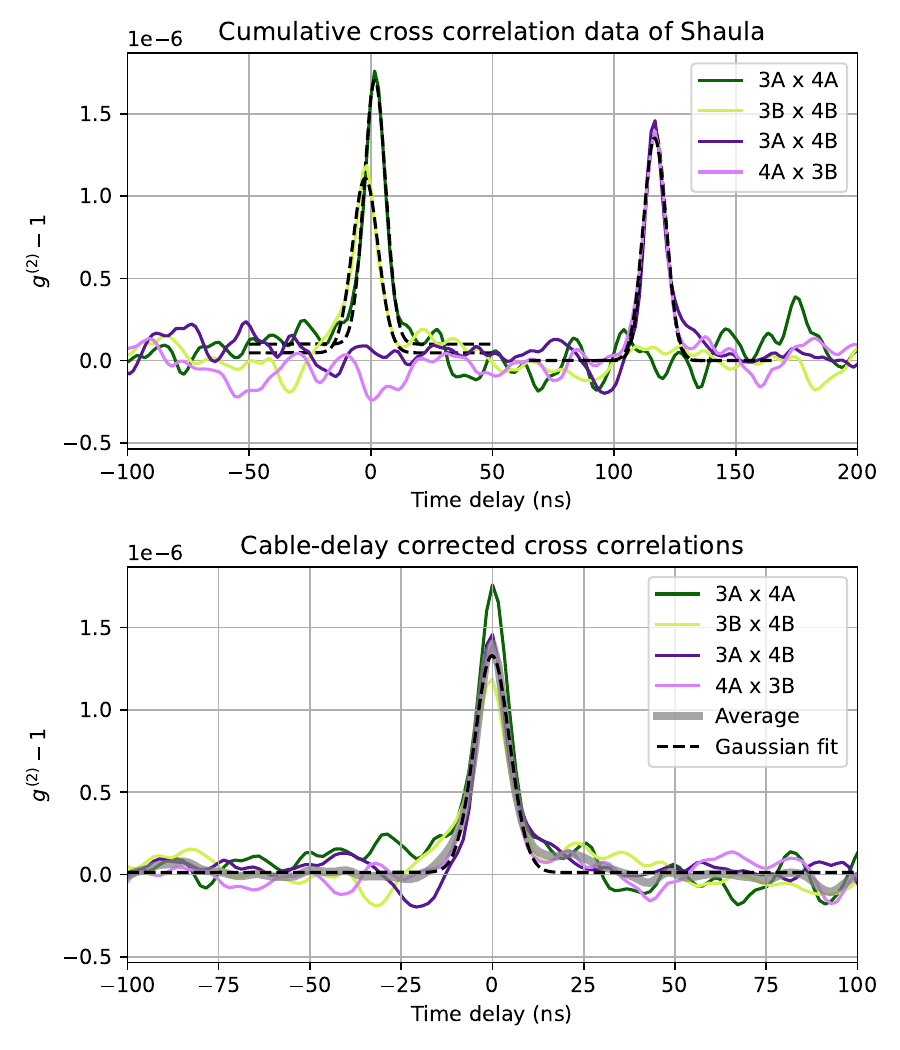}
    \caption{Top: bunching peaks of the four cross correlation channels measured with Shaula, accumulated over $16\,$hours. Bottom: the four peaks are shifted to zero time delay and averaged.}
    \label{fig:avg}
\end{figure}

Fig.~\ref{fig:avg} (upper) shows the four cross correlation channels of the star Shaula by accumulating data over the entire measurement time ($\approx 19\,$h). By doing this one averages over all the different spatial coherences present during the measurement period and thus cannot extract specific spatial coherence information. Instead, due to the large statistics the bunching signals are of high significance and can be used to determine the exact time delay for each correlation channel. In Fig.~\ref{fig:avg} (upper) the two time delay regions around $0$ and $117\,$ns are visible among the four correlations. By applying Gaussian fits to the bunching peaks, the cable delay times can be determined with low uncertainty. It is also apparent that with increasing total cable lengths the signal height (and also integral) decreases, resulting in the highest signal for the correlation of both $10-$m channels, and the smallest signal for both $40-$m channels. This behaviour is not expected and under investigation in the lab for future campaigns, but appears to be consistent amongst the different observed targets and thus doesn't falsify later evaluations. Effects of potential changes in cable lengths due to thermal expansions are on the order of less than $100\,$ps signal time delay and can be neglected.

\subsection{Data averaging and parameter fixing}
Since each of the four cross-correlation channels is an independent measurement, it should be possible to independently use every one of them to observe the change in spatial coherence with changing projected telescope baselines, and from that extract the angular size of the star. However, we recognize that upon dividing the complete measurement time of a target into reasonably short time segments for investigation of the change in spatial coherence (for a detailed explanation of what is called reasonable and how it is done see section \ref{sec:chunks}), the signal-to-noise of the bunching peaks - especially at large baselines and thus low spatial coherence - is not high enough to conveniently fit a Gaussian function. Therefore we use all four correlation channels in each time segment and average over them. In Fig.~\ref{fig:avg} (lower) this is at first done for the the total accumulated time of Shaula. With the determined delay times from the upper plot each of the channels is shifted in time to overlay the bunching peaks at zero time delay, and then averaged inversely proportional to the square of the standard error of the data points excluding the peak region. The averaged signal then has smaller fluctuations and thus allows for improved fitting of Gaussian functions to the bunching peaks even at shorter measurement times.
However, we recognize that it is beneficial to fix the mean and the width of the Gaussian function when fitting into a single segment of the data, and only fit the amplitude. The mean and the width used for any segment analysis are extracted from a Gaussian fit to the time-accumulated bunching peak in Fig.~\ref{fig:avg} (lower).

\subsection{Measurement segments}\label{sec:chunks}
For a detailed spatial coherence analysis the accumulated data has to be divided into smaller time segments in which the correlation is calculated. The choice of the segment time length is a trade off between uncertainties in horizontal and vertical direction in the spatial correlation plots of coherence time versus baseline (see e.g. upper right plot in Fig.~\ref{fig:shaula_sc}): the smaller the time segment is chosen to be, the less projected baseline range is being scanned by the telescopes during that time, and hence the smaller the uncertainty in the baseline. However, short measurement times result in less photon statistics, the signal-to-noise of the bunching peaks is smaller and the uncertainties in determining the coherence times become larger. Latter effect also depends on the photon rates, bright stars allow for shorter measurement segments, whereas fainter stars require longer ones.

\begin{table}
\centering
\begin{tabular}{| c | c | c | c |} 
 \hline
 Evening & Shaula & Nunki & Acrux \\ [0.5ex] 
 \hline\hline 
 04/16 & -                     & -                     & $ 5 \times 29.78\,$min \\
 04/17 & -                     & -                     & $15 \times 30.52\,$min \\
 04/18 & -                     & $3 \times 87.45\,$min & $ 4 \times 26.34\,$min \\
 04/19 & $4 \times 91.84\,$min & -                     & -                      \\
 04/20 & $4 \times 68.49\,$min & -                     & -                      \\
 04/21 & $4 \times 79.78\,$min & -                     & -                      \\
 04/22 & -                     & $3 \times 74.91\,$min & -                      \\
 04/23 & -                     & $2 \times 63.68\,$min & -                      \\
 \hline
 \textbf{Total} & 16h 0min & 10h 14min & 11h 52min \\
 \hline
\end{tabular}
\caption{Overview of the measurement segments and segment times for the three main targets of the campaign, Shaula, Nunki and Acrux.}
\label{table:measurement_chunks}
\end{table}

Table \ref{table:measurement_chunks} lists the separation into measurement segments for Shaula, Nunki and Acrux in each observation night. Since the photon rates of the bright star system Acrux ($0.76\,$mag, \cite{1966MNSSA..25...44C}) are significantly ($\approx$ factor 2) higher than for Shaula and Nunki, which both are similarly bright, we divided the Acrux data into smaller measurement segments. As the total measurement time of a target in different observation nights vary, we also used slightly different segment times for the same target in different nights in order to optimally use the observation time of a night without removing a significant amount of data that didn't fill up to a segment.

\subsection{Spatial coherence curves} \label{results}
\begin{figure*}
    \includegraphics[width=0.99\linewidth]{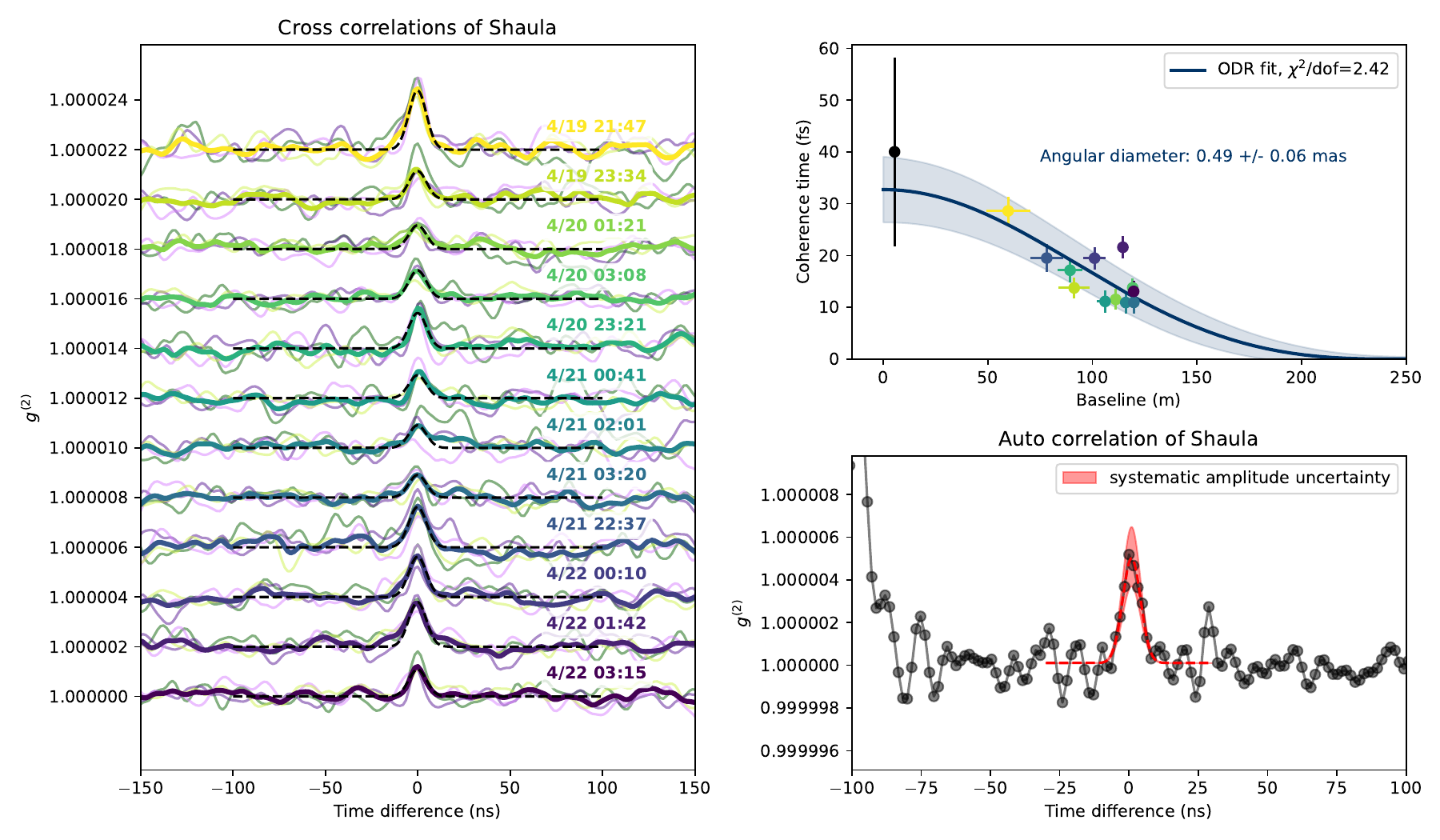}
    \caption{$g^{(2)}$ cross correlation functions of each measurement segment (left), the zero-baseline correlation (bottom right) and the corresponding spatial correlations (top right) of Shaula}
    \label{fig:shaula_sc}
\end{figure*}

Fig.~\ref{fig:shaula_sc} (left) shows the $g^{(2)}$ functions of the different measurement segments of Shaula. For a better visualization of single measurements, they are separated from each other in vertical direction. The correlation of the four single channels are displayed as thin lines with the colour code being the same as in Fig.~\ref{fig:avg}. The thick lines are the averages of theses channels, which are used to apply Gaussian fits (black dashed lines) with the amplitude being the only fit parameter. Mean and width are extracted from the time-averaged signal in the lower plot in Fig.~\ref{fig:avg}, we do not fit an offset to the correlations. The integral of the fit is the coherence time as defined in section \ref{sec:observables}. The displayed time stamps are UTC and denote the centre time of each measurement segment.

For each measurement segment we also calculate the corresponding telescope baseline value. Each $3.4$-second measurement file of a segment can be mapped to a corresponding baseline. We average over all these baseline values within a segment in order to get a single value, the $1\sigma$ fluctuation of all the baselines in a segment is used as uncertainty on the baseline (horizontal error).

\begin{figure*}
    \includegraphics[width=0.99\linewidth]{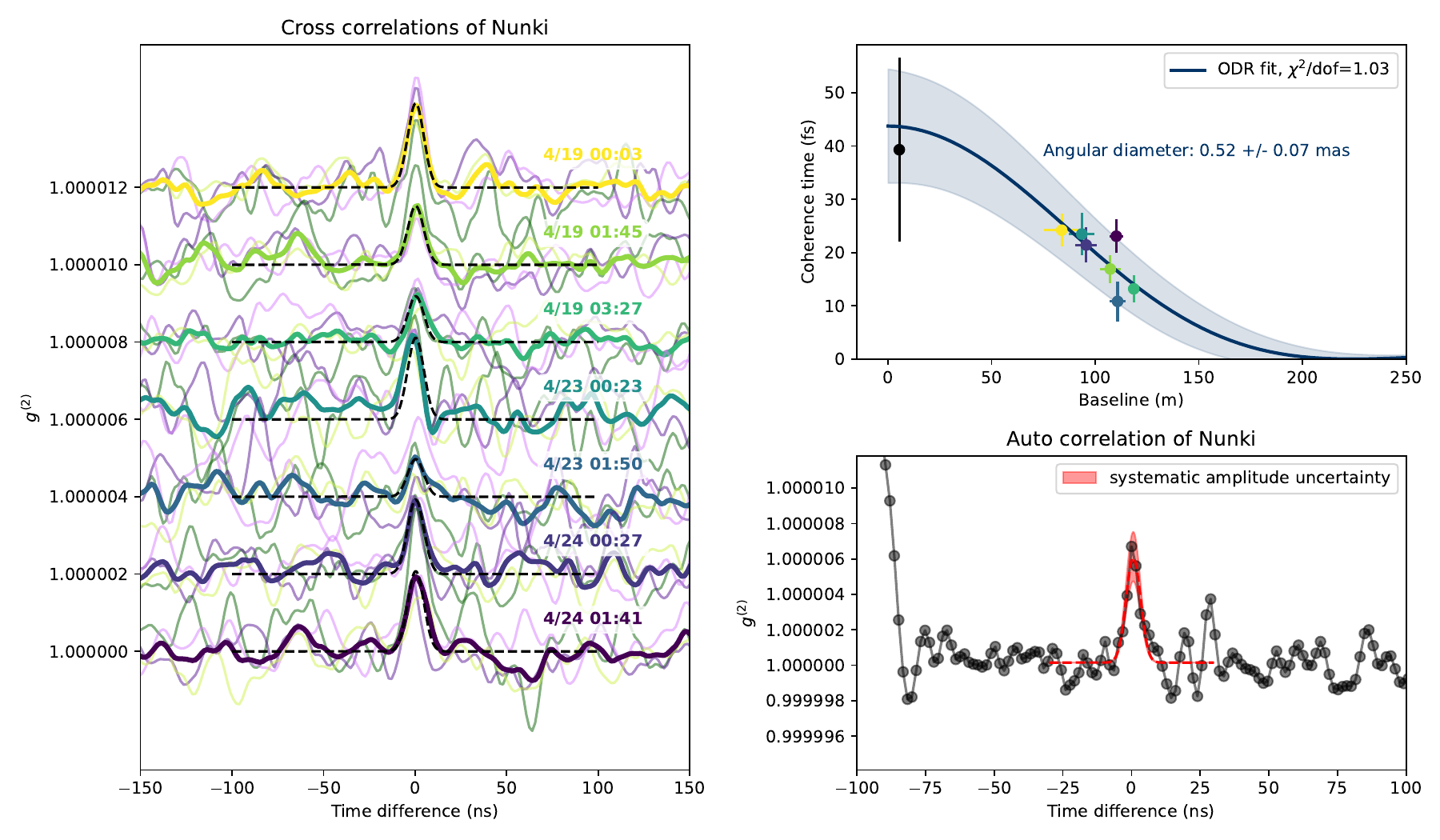}
    \caption{$g^{(2)}$ cross correlation functions of each measurement segment (left), the zero-baseline correlation (bottom right) and the corresponding spatial correlations (top right) of Nunki}
    \label{fig:nunki_sc}
\end{figure*}

In Fig.~\ref{fig:shaula_sc} (top right) the coherence times are plotted versus the corresponding average baseline of the measurement segment. The black data point close to zero baseline is the auto correlation averaged for both telescopes over the entire measurement time. Even though it is called zero-baseline correlation before, we plotted it at $(5.4 \pm 2.5)\,$m as this is the mean distance of two photon impact positions on the mirror of one telescope. The auto correlation is plotted in the lower right plot. As mentioned before, it is expected to have the highest signal-to-noise and smallest uncertainty of the coherence time due to accumulating statistics over the whole measurement time. However, since we find systematic noise contributions in both telescopes affecting the auto (but not the cross) correlations, the auto correlation suffers from a high uncertainty. To estimate the effect of systematic oscillations in the baseline, we take the strongest up- and down oscillations measured in the auto-correlation of Acrux (see section \ref{sec:zero_baseline}), averaged over the entire measurement time, in the time range displayed in fig.~\ref{fig:acrux_ac}. These fluctuations add as an uncertainty to the measured amplitude, and therefore to the peak integral. The systematic uncertainties are displayed as red band around the fit in the auto-correlations (figures \ref{fig:nunki_sc}) and \ref{fig:shaula_sc}.

Fig.~\ref{fig:nunki_sc} shows cross correlations, auto correlations and the spatial coherence values for Nunki.

\subsection{Data analysis}\label{sec:data_analysis}
For our model we assume that Nunki can be treated as a single star, since the primary component is dominant in the system and potential companions do not significantly contribute to the measured photon rates. It is an appropriate approximation, since Nunki only has a mag 9.95 companion \citep{mason20012001}. We use a uniform disc model and fit the corresponding function to the spatial coherence data of Nunki \citep{davis2000limb}:
\begin{equation}\label{eq:airy}
    f(x) = a \left[ \frac{2J_1(\pi x \varphi/\lambda)}{\pi x \varphi/\lambda} \right]^2
\end{equation}
with $x$ being the baseline, $\lambda = 465\,$nm the operating wavelength and $J_1$ being the Bessel function of first kind. The two fit parameters are the constant $a$ and the angular diameter $\varphi$.

The constant $a$ corresponds to the coherence time at zero baseline and should thus be equal to Eq.~\ref{eq:coherence_time} with $k_s = 1$, resulting in $a_\textnormal{exp} = 152\,$fs. We however obtain $a_\textnormal{Nunki} = (43.76 \pm 10.40)\,$fs, indicating a coherence loss factor of about $3.5$. The reasons for this loss are yet unclear. We are investigating potential coherence losses due to the use of long signal cables, as is already indicated in Fig.~\ref{fig:avg}, where correlations between channels with long cables are significantly smaller than correlations between channels with short cables. We further identify the narrow bandpass filters as potential source of coherence loss. If light on these interference filters comes in at an angle, the transmitted wavelengths are shifted to smaller values. Light with different inclination angles on the filter therefore broadens the transmission spectrum of the filters and results in reduced temporal coherence. For that reason future intensity interferometry observations at H.E.S.S. will feature broader wavelength filters of $10\,$nm FWHM (see also section \ref{sec:conclusion}).

The second fit parameter $\varphi$ directly yields the angular diameter of the star, which is main subject of observation. We obtain an angular diameter of $\varphi_\textnormal{Nunki} = (0.52\pm0.07)\,$mas.

\begin{figure}
    \includegraphics[width=\linewidth]{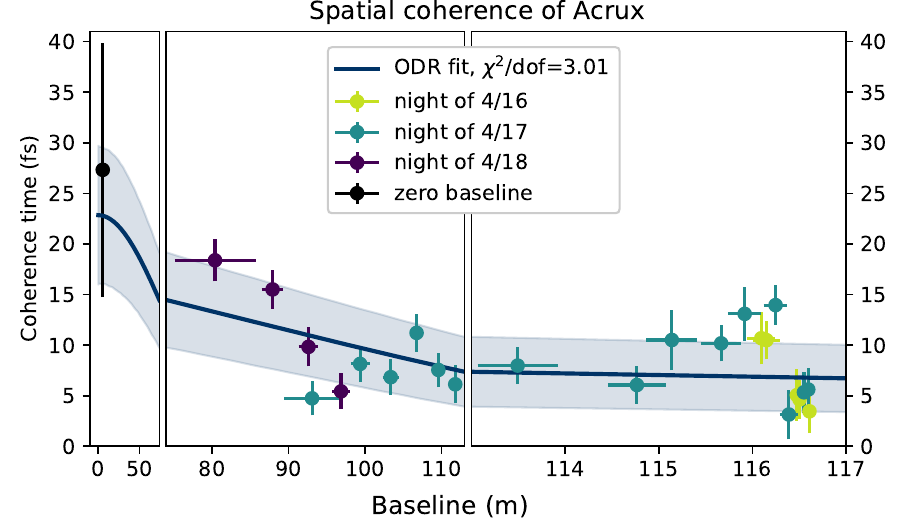}
    \caption{Spatial correlations of Acrux. The horizontal axis is divided into three parts for better visualization. The colours mark the three measurements nights of Acrux. The zero-baseline data point is marked in black on the very left of the plot.}
    \label{fig:acrux_sc}
\end{figure}

Treating the other two observed targets as single stars is not suitable. Shaula is a triple star system \citep{handler2013time}, with the two main components having a magnitude ratio of $0.66 \pm 0.10$ \citep{tango2006orbital}. The constellation of Acrux consists of many individual components - ($\alpha^1$ Cru and $\alpha^2$ Cru can even be separated visually) \citep{kolaczek2020alpha}. In both cases we do not expect the data to (fully) agree with a single star fit model. We however applied a fit according to equation \ref{eq:airy} (see figures~\ref{fig:shaula_sc} and \ref{fig:acrux_sc}), and conclude from the fact that the reduced $\chi^2$ for Shaula ($2.42$) and Acrux ($3.01$) are significantly larger than for Nunki ($1.03$) that the data in fact don't align well with a single star model. Especially for Acrux we took many data at large baselines, which show indications of a more complex source geometry. Potential future observations may allow for more detailed analyses treating these systems as binary or multiple star systems and extracting both orbital and stellar parameters.

Despite Shaula being treated best as a binary (neglecting the faint secondary component) \citep{tango2006orbital}, the single star fit model does contain information about the angular diameter of the single stars. Even though the fit misses fast oscillations in the spatial coherence curve, induced by the angular separation of both dominant stars, the overall trend of the curve is related to the angular diameters of the stellar disks. Assuming both stars have similar angular diameters, we find $\varphi_\textnormal{Shaula} = (0.49 \pm 0.06)\,$mas for both stars. This value, however, should be treated with care. The assumption of similar stellar radii as well as the arbitrary sampling of the potentially fast oscillating spatial coherence curve are most likely causing systematic errors.

\subsection{Data comparison}
From literature one finds $\varphi_\textnormal{Nunki} = 0.68\,$mas \citep{underhill1979effective}, computed from photometry data \citep{blackwell1977stellar}, which shows some deviance to our measurements of $(0.51\pm0.05)\,$mas. For Shaula we find $(5.2 \pm 1.7)\,\textup{R}_\odot$ for the primary and $(5.3 \pm 1.3)\,\textup{R}_\odot$ for the tertiary star \citep{handler2013time}, which supports our assumption of similar radii in section~\ref{sec:data_analysis}, and from the Hipparcos parallax results a distance of $d_\textnormal{Shaula} = (175 \pm 23)\,$pc. However, \cite{tango2006orbital} obtain a distance of $d_\textnormal{Shaula} = (112 \pm 5)\,$pc. Using the latter we calculate an angular diameter of $\varphi_\textnormal{Shaula} = (0.43 \pm 0.14)\,$mas, whereas \citep{underhill1979effective} obtain an angular diameter of $0.64\,$mas computed from infrared photometry data \citep{blackwell1977stellar}. We make clear, that due to the disagreeing set of parameters that is found in literature, as well as our simplified description model of Shaula, we are not making a strong statement about an agreement of our measurements with previous ones.\\

\section{Conclusion and outlook}\label{sec:conclusion}

We equipped two of the Phase I telescopes of the H.E.S.S. gamma ray observatory with an external intensity interferometry setup for the time of the moonlight measurement break in April 2022. We measured the spatial coherence curves of two stars and were able to extract angular diameters with a precision at the $10\,$per cent scale, while combining auto-correlation measurements in the same telescope with cross-correlation measurements between two telescopes. We noticed that the absolute measured coherence times are smaller than expected, which could partly be caused by imperfections in the light inclination at the small ($2\,$nm) bandwidth interference filters, which limits signal-to-noise of the observations.

As for the next measurement campaign in April/May 2023 some changes are made to the setup.\\
First of all the bandwidth of the interference filters are broadened to $10\,$nm instead of $2\,$nm. As described in \autoref{sec:data_analysis} this results in more stable optics, and therefore potentially less spatial coherence losses and higher signal-to-noise.\\
Furthermore, we are expanding our setup to two colour measurements. Each of the two PMTs on one telescope has their own band-pass filter at a different wavelength ($375$ and $470\,$nm) making it possible to measure with two colours simultaneously - at the cost of the zero baseline measurement. Both measurement modes are interchangeable however.\\
The third adjustment is equipping a third telescope with our setup to gain more telescope baselines.

\section*{Acknowledgements}
We thank the H.E.S.S. collaboration for reviewing this work and for allowing us to use their telescopes for our measurement campaign.
We also thank the H.E.S.S. local crew for their support on site and the operations team, especially Dmitriy Kostunin and Gianluca Giavitto, for their help online and with installing our setup. 
Thanks to Dmitry Malyshev for his contributions to preparations and discussions, and to Alison Mitchell for her thoughts and input on this work.
This work was supported with a grant by the Deutsche Forschungsgemeinschaft (\lq Optical intensity interferometry with the H.E.S.S. gamma-ray telescopes\rq - FU 1093/3-1).

\section*{Data Availability}
The data underlying this article will be shared on reasonable request to the corresponding author. Correlation histograms are available in time-intervals of $3.436\,$s. Due to the large size of the digitized waveforms in excess of several TB, the raw data can not be made available online.



\bibliographystyle{mnras}
\bibliography{cites} 








\bsp	
\label{lastpage}
\end{document}